\documentclass[conference]{IEEEtran}
\usepackage[dvips]{graphicx}
\usepackage{amsfonts,amssymb}
\usepackage{cite}
\usepackage{epsfig}
\usepackage{makeidx}
\usepackage{color}

\begin{document}
\title{Negative Selection Approach Application in Network Intrusion Detection Systems}
\author{
\IEEEauthorblockN{Amira Sayed A. Aziz}
\IEEEauthorblockA{Universite Francaise d'Egypte (UFE)\\
Cairo, Egypt\\
Scientific Research Group in Egypt (SRGE)\\
Email: amiraabdelaziz@gmail.com \\
\\
\IEEEauthorblockN{Aboul ella Hassanien}
Head of Scientific Research Group in Egypt (SRGE)\\
Faculty of Computers \& Information - Cairo University \\
Cairo, Egypt}
\and
\IEEEauthorblockN{Ahmad Taher Azar}
\IEEEauthorblockA{Faculty of Computers \& Information, Benha University\\
Benha, Egypt \\
Scientific Research Group in Egypt (SRGE)\\
Email: ahmad\_t\_azar@ieee.org \\
\\
\IEEEauthorblockN{Sanaa El-Ola Hanafy}
Faculty of Computers \& Information - Cairo University \\
Cairo, Egypt}}

\maketitle
\begin{abstract}
Nature has always been an inspiration to researchers with its diversity and robustness of its systems, and Artificial Immune Systems are one of them. Many algorithms were inspired by ongoing discoveries of biological immune systems techniques and approaches. One of the basic and most common approach is the Negative Selection Approach, which is simple and easy to implement. It was applied in many fields, but mostly in anomaly detection for the similarity of its basic idea. In this paper, a review is given on the application of negative selection approach in network security, specifically the intrusion detection system. As the work in this field is limited, we need to understand what the challenges of this approach are. Recommendations are given by the end of the paper for future work.
\end{abstract}

\section{Introduction}
Networks are more vulnerable by time to intrusions and attacks, from inside and outside. Cyber-attacks are making news headlines worldwide, as threats to networks are getting bolder and more sophisticated. Reports of 2011 and 2012 are showing an increase in network attacks, with Denial of Service (DoS) and targeted attacks having a big share in it. As reported by many web sites like \cite{l}\cite{m}\cite{n}, figures 1 and 2 show motivations behind attacks and targeted customer types respectively.

\begin{figure}[h]
\centering
{\includegraphics[width=3.0in,height=2.4in]{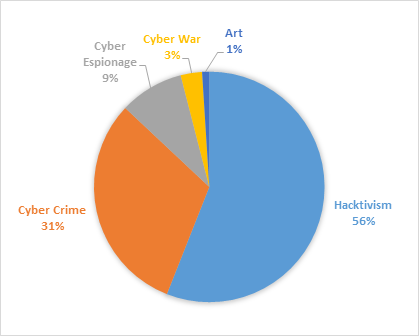}}
\caption{Motivation behind attacks.}
\end{figure}
\begin{figure}[h]
\centering
{\includegraphics[width=3.0in,height=2.6in]{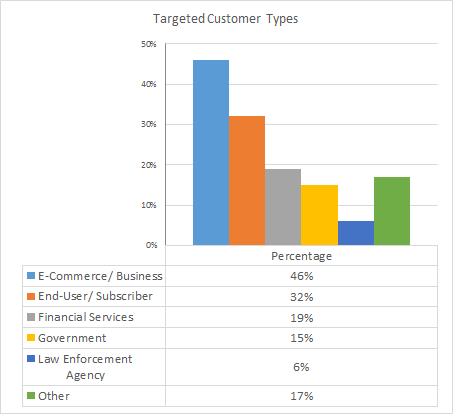}}
\caption{Internal Network Security Concerns.}
\end{figure}
Internal threats and Advanced Persistent Threats (APT) are the biggest threats to a network, as they are carefully constructed and dangerous, due to internal users' privileges to access network resources. Figure 3 shows internal network security concerns. With this in mind, and the increasing sophistication of attacks, new approaches to protect the network resources are always under investigation, and the one that is concerned with inside and outside threats is the Intrusion Detection System. 

\begin{figure}[h]
\centering
{\includegraphics[width=3.0in,height=2.8in]{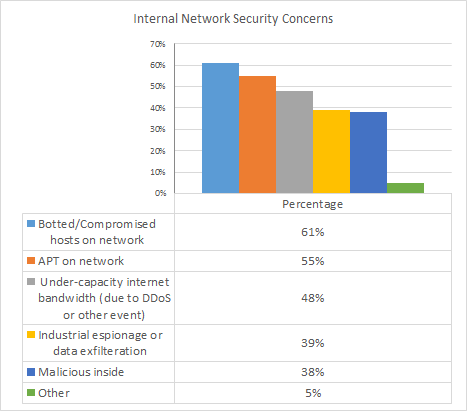}}
\caption{Internal Network Security Concerns.}
\end{figure}
Intrusion detection systems \cite{39}\cite{40}\cite{41} have been around for quite some time, as a successful security system. An Intrusion Detection System (IDS) is a system that defines and detects possible threats within a computer or a network, by gathering and analysing information from the surrounding environment. An IDS can be classified – based on detection methodology – as anomaly-based and misuse-based. An anomaly IDS compares monitored activities to a normal model that it built earlier using the analysed information that is been collected from the system. A misuse IDS is signature based, which means that it detects anomalies by comparing the monitored activities to a database of attack patterns that was built based on the analysed information.

Intrusion detection systems can also be classified based on the system it protects, to a Network-based IDS (NIDS) and Host-based IDS (HIDS). A NIDS monitors a network communication activities while a HIDS monitors a single host’s activities through its audit logs to detect anomalies. Many Computational Intelligence (CI) approaches were applied in IDSs implementation, one of them is artificial immune systems.

Artificial Immune Systems (AIS) \cite{1}\cite{2}\cite{3} are a set of methodologies inspired by the Human Immune System (HIS), and are considered a branch of computational intelligence bio-inspired technology. AIS connects immunology with computer science and engineering. Attention was drawn to immune system as an inspiration to new approaches to solve complex problems. It mimics the HIS which is adaptive, distributed, tolerant, self-protective, and self-organizing with its many naturally-embedded techniques such as learning, feature extraction, and pattern recognition. There are many methodologies within the immune system that can form an inspiration to a wide range of techniques. Those methodologies are Negative Selection Approach (NSA) \cite{5}, Artificial Immune Networks (AIN) \cite{7}\cite{8}, and Clonal Selection Algorithm (CSA) \cite{9}. Recent theories have also emerged such as Danger Theory (DT) \cite{11}\cite{12}, Dendritic Cells Algorithms (DCA) \cite{13} and Pattern Recognition Receptor Model (PRRM) \cite{15}\cite{16}.

Negative Selection Algorithms are based on the concept that an immune system discriminates between self and non-self cells, and consider non-self as intruders to the body/system. This is held by the T-cells, which are originally created in the bone marrow. Then, they are moved to the thymus for the maturation process, where the T-cells learn the self (normal) patterns for the negative selection process. Those that are activated by self antigens are destroyed, until the maturation process is complete to those which are mature enough to match and mark the non-self antigens. Finally, mature T-cells are released to the system to start the detection process. In IDSs, a model is built to represent the normal behaviour or pattern of a system, where the generated detectors are trained using that model. Then these detectors are released to the system where they negatively detect and define anomalous activities, as shown in figure 4. 
\begin{figure}[h]
\centering
{\includegraphics[width=2.2in,height=4in]{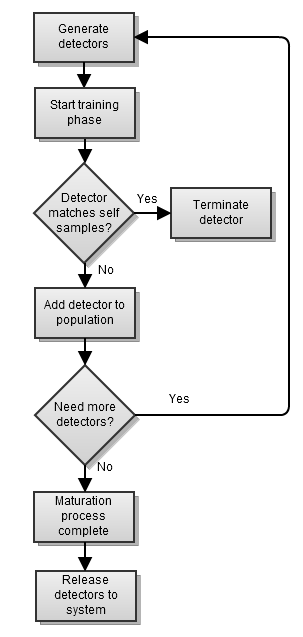}}
\caption{IDS detectors generation process.}
\end{figure}
The rest of the paper is organized as: Section II presents related reviews published before, Section III presents a comparative analysis between different work applied using NSA, and finally Section IV is about a discussion and the conclusions of that discussion.

\section{Related Work}
There was not any previous specific reviews done on the NSA application, neither in network security field nor any field in general. Multiple reviews and studies have been done to follow the evolution of AIS methodologies and their application to different fields, which are introduced in this section.

Gonzalez submitted a PhD thesis \cite{17} in 2002, which was investigating into NSA and proposed new detector generating algorithms for different representation schemes of NSA. In 2003 Dasgupta et al. \cite{18} gave a review over AIS techniques and researches made from 1999 to 2003 on AIS work and applications. Timmis discussed in \cite{19} the challenges AIS applications may face, to take into consideration. He concluded the following challenges of AISs as (1) the need of more interaction with immunologists and mathematicians for the creation of useful models through experimentation, (2) theoretical and formal basis for AIS is required to understand the nature of AIS and the best and more fitting application of it, and (3) more interaction between immune systems with  other systems is essential and more attention should be paid to integrations for better functioning. 

Timmis also gave a review on the current (back then) state of AIS approaches \cite{20} and listed more challenges for future development in this area. In that same year (2006), Dasgupta \cite{21} presented a study on AIS components and functions inspired by biological immune system different functional elements. It also provided a time-line of recent AIS developments, focusing on the Computer Security and Fault Detection areas of applications. He showed the process of how to apply an immune algorithm to solve a problem, and it is shown in figure 5.
\begin{figure}[h]
\centering
{\includegraphics[width=1.7in,height=2.1in]{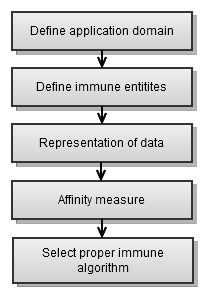}}
\caption{Solving a problem using an AIS.}
\end{figure}
Multiple studies and reviews were published in 2007. Two of them \cite{22}\cite{23} were bibliographies of all thesis studies, publications, and systems developed in the field of AIS. Hart and Timmis \cite{24} gave a look into the contributions of the AIS methodologies, and what they brought to different application areas, such as: Clustering and Classification, Anomaly Detection, Computer Security, Optimization, and Learning.

In \cite{25}, a review of different AIS approaches to intrusion detection systems along with the implementations of different algorithms and their results were given. They discuss it from the point of view that most of the techniques followed to build an IDS are not able to cope with the dynamic and complex nature of computer systems security. In 2009, Garcia et al. \cite{26} presented a literature review on the recent years work of malicious activity detection methods using AIS, and the available platforms and research projects in this area. In \cite{27}, they presented a review of CI core methods and their application in intrusion detection. The most applied CI approaches in intrusion detection are: Artificial Neural Networks (ANN) \cite{45}\cite{46}, Fuzzy sets \cite{47}\cite{48}, Evolutionary Computation (EC) \cite{49}, AIS, and Swarm Intelligence (SI) \cite{51}.  In 2010, multiple studies were made \cite{3}\cite{28}\cite{30} giving a review on recent work and advances in AIS and their applications.

\section{Comparative Analysis}
Forrest et al. \cite{5} were the first ones to come up with the idea of using NSA for anomaly detection. Since then, researches have been coming up with studies and developing models inspired by NSAs. Most of them were about learning mechanisms to create rules (detectors) that can be used to match patterns of self and non-self. The number of works discussing the NSA from the point of view of network security are very limited compared to applying NSA in other areas and applying AIS in general. In this section, a list of published researches using NSA in NIDS are presented.

Gonzalez and Dasgupta cooperated in a number of papers \cite{a}\cite{b}\cite{c} where they used Genetic Algorithms (GA) to generate a set of self detectors based on NSA concept. They tested their work on 3 real-valued features from the popular data set then –-- the DARPA generated traffic data \cite{d} –-- and their work gave very good results. They also implemented a system where positive selection (where detectors can detect anomalous activities based on non-self patterns) was applied instead of negative selection, but the results showed that negative selection gives better detection results than the positive selection which is mostly applied in the IDSs.

In \cite{31}, they built a distributed IDS using NSA to obtain set of detectors which exchange status information through P2P connections. Data Set used was a network simulation with simulated attack scenarios. The collaboration between the AIS clients decreases the FPR to 18\% and 10\% on two different feature vectors. Powers and He \cite{32} applied NSA to generate detectors with objectives that increase the generality of detectors with low false alarms rate. The Self Organizing Maps (SOM) were integrated in the system as a second phase to classify the detected anomalies. They used the KDD Cup’99 data set \cite{e} for testing, and detection rates were: 99.4\% for Normal, 96.8\% for DoS, 64.7\% for Probe, 34.6\% for U2R, and 5.2\% for R2L. NSA in combination with Decision Trees (DT) was applied for comparison purposes in \cite{33} to DCA. The algorithm was tested using the KDD Cup’99 data set and the average True Positives Rate (TPR) was 74.17\% and average False Positives Rate (FPR) was 0.005\%. In \cite{34} they implemented an antigen feedback mechanism to provide an efficient way for detectors generation in a short period of time. They tested their system using the KDD Cup’99 dataset and they achieved a DR of 95.21\% on attack strings and a DR of 99.21\% on normal strings. They also produced the project Arisytis (Artificial Immune Systems Toolkits) as a project that can be used by any researcher for further work. 

NSA was combined with Rough Sets for optimized feature selection in \cite{36} and tested on KDD Cup’99 data set and achieved a TPR up to 98.25\% and a True Negatives Rate (TNR) up to 99.97\%. V-detectors (variable-size detectors) were generated in \cite{37} based on ideas from NSA combined with Restricted Coulomb Energy (RCE) neural networks, which are designed specifically for hyper-sphere classifiers. The test ran on the NSL-KDD data set \cite{f} in 10 trials, with number of detectors between 600 and 625. Average DR achieved was 99.88\%, average false alarms was 4.51\%, and average accuracy was 90.2\%. In \cite{38}, they proposed an algorithm FtNSA (Further training NSA), where a strategy is adapted to generate self-detectors that cover the self region in a way that reduces the self samples for the testing phase. They tested their system on 7 data sets, including the KDD Cup’99 data set. Compared to NSA, the FtNSA has a little lower DR but it also has reduced FPR while maintaining a stable performance. FtNSA DR was between 95 and 96.5\% versus 96.5 to 98\% for NSA, and the FPR of FtNSA was slightly around 2\% versus 4 to 16\% using NSA.

\section{Conclusion \& Discussion}
NSA has been very popular to research and apply in different areas, as it is simple and easy to implement. If detectors are well generated, then the detection process can come up with very good results. A good detector should not be matching normal patterns as harmful particles, so the selection of an algorithm to generate such detectors is very important. The affinity measures – that are used for the matching rules – should be accurately selected, depending on whether self components should be exactly the same or a degree of similarity is involved.

Through the review of the work accomplished so far to apply NSA in network security, we can realize that limited work was done compared to applying NSA in other areas. The NSA algorithm is the most compatible to anomaly detection approach, as they share the same process of the discrimination between normal and anomalous activities and components in the system. Through the reviewed work, we can come up with following notes:
\begin{itemize}
  \item NSA should be combined with other classification methods (such as decision tree and neural networks), as NSA basically classify activities to normal and anomaly only. 
  \item The generation process of the detectors should not be random for the detectors to be effective. The generation process should also happen in a reasonable time.
  \item The definition of the matching rules is very important, so that the NSA application would not generate high false alarms.
  \item One of the most difficult challenge to NSA is that self elements do not remain the same through the whole time, and they may change from time to time. Continuous learning is a basic need in NSA so that detectors adjust themselves through time to be compatible with the self components representation. 
  \item Communication between detectors is also important to update their rules from time to time with new information.
\end{itemize}

\end{document}